# Transmission Strategies in Multiple Access Fading Channels with Statistical QoS Constraints

Deli Qiao, Mustafa Cenk Gursoy, and Senem Velipasalar

**Abstract**

Effective capacity, which provides the maximum constant arrival rate that a given service process can support while satisfying statistical delay constraints, is analyzed in a multiuser scenario. In particular, the effective capacity region of fading multiple access channels (MAC) in the presence of quality of service (QoS) constraints is studied. Perfect channel side information (CSI) is assumed to be available at both the transmitters and the receiver. It is initially assumed the transmitters send the information at a fixed power level and hence do not employ power control policies. Under this assumption, the performance achieved by superposition coding with successive decoding techniques is investigated. It is shown that varying the decoding order with respect to the channel states can significantly increase the achievable throughput region. In the two-user case, the optimal decoding strategy is determined for the scenario in which the users have the same QoS constraints. The performance of orthogonal transmission strategies is also analyzed. It is shown that for certain QoS constraints, time-division multiple-access (TDMA) can achieve better performance than superposition coding if fixed successive decoding order is used at the receiver side.

In the subsequent analysis, power control policies are incorporated into the transmission strategies. The optimal power allocation policies for any fixed decoding order over all channel states are identified. For a given variable decoding order strategy, the conditions that the optimal power control policies must satisfy are determined, and an algorithm that can be used to compute these optimal policies is provided.

## I. INTRODUCTION

In wireless networks, the design and analysis of efficient transmissions strategies have been of significant interest for many years. In particular, fading multiple access channels (MAC) have been extensively studied

The authors are with the Department of Electrical Engineering, University of Nebraska-Lincoln, Lincoln, NE, 68588 (e-mails: dqiao726@huskers.unl.edu, gursoy@engr.unl.edu, velipasa@engr.unl.edu).

This work was supported by the National Science Foundation under Grants CCF – 0546384 (CAREER), CNS – 0834753, and CCF – 0917265. The material in this paper was presented in part at the the IEEE International Conference on Communications (ICC), Cape Town, South Africa, in May 2010.

from an information-theoretic point of view [1]-[8]. For instance, Tse and Hanly [4] have characterized the capacity region of and determined the optimal resource allocation policies for multiple access fading channels. They have shown that the boundary surface points are in general achieved by superposition coding and successive decoding techniques, and obtaining each boundary point can be associated with an optimization problem in which a weighted sum rate is maximized. Vishawanath *et al.* [7] derived the explicit optimal power and rate allocation schemes (similar to *waterfilling*) by considering that the users are successively decoded in the same order for all channel states. For the convex capacity region, the unique decoding order was shown to be the reverse order of the priority weight. While superposition coding and successive decoding strategies provide superior performance, time-division multiple access (TDMA) may in certain cases be preferred due to its simplicity. Note that the performance of TDMA approaches that of the optimal strategy as the signal-to-noise ratio (SNR) vanishes but, as shown by Caire *et al.* in [8], TDMA is strictly suboptimal when SNR is low but nonzero.

While establishing the fundamental performance limits, the above-mentioned studies have not explicitly taken into account buffer constraints and random arrivals. In [9] and [10], Yeh and Cohen considered multiaccess fading channels with random packet arrivals to buffered transmitters, and characterized rate and power allocation strategies that maximize the stable throughput of the system. In [11], the same authors investigated rate allocation policies that minimize the average packet delay in multiaccess fading channels again under the assumption of randomly arriving packets.

In this paper, we also investigate the performance under buffer constraints but provide a perspective different from those of previous studies. In particular, we consider statistical quality of service (QoS) constraints in the form of limitations on the buffer violation probabilities, and study the achievable rate region under such constraints in multiaccess fading channels. Note that in certain delay sensitive applications, such as interactive or streaming video, constraints on delay bound violation probability may be required rather than limitations on the average delay. For this analysis, we employ the concept of effective capacity [12], which can be seen as the maximum constant arrival rate that a given time-varying service process can support while satisfying statistical QoS guarantees. Effective capacity formulation uses the large deviations theory and incorporates the statistical QoS constraints by capturing the rate of decay of the buffer occupancy probability for large queue lengths. The analysis and application of effective capacity in various settings has attracted much interest recently (see e.g., [13]–[16] and references therein). In [16], Liu *et al.* considered a two-user cooperative multiple access fading channel and analyzed the rate region achieved with frequency-division multiplexing when the users are operating under QoS constraints in the form of limitations on buffer overflow probabilities. In this study, cooperation among the users is shown to significantly improve the achievable rate region if the quality of the wireless link between the users is better than those of the links between the users and the destination. We note that since the transmitters are assumed to not know the channel conditions, power and rate adaptation policies are not studied in [16]. Additionally, since orthogonal transmission schemes are considered, superposition coding and successive decoding strategies are not addressed in detail.

Our contributions and major findings in this paper can be summarized as follows. We consider the scenario in which both the transmitters and the receiver have perfect channel side information (CSI). First, assuming that no power control is employed in the transmission, we characterize the rate regions for both superposition transmission strategies and TDMA. Unlike the results obtained in [1] and [7], varying the decoding order with respect to the channel states is shown to significantly increase the achievable rate region (i.e., *throughput region*) under QoS constraints. Also, it is demonstrated that time sharing strategies among the vertex of the rate regions can no longer achieve the boundary surface of the throughput region. Additionally, we show that if we take the sum-rate throughput, or the sum effective capacity, as the performance metric, TDMA can in certain cases even achieve better performance than superposition coding when fixed decoding order is employed at the receiver. Next, we incorporate power control policies into the model. For this case, we first obtain closed-form expressions for the optimal power control policies under the assumption that the decoding order is fixed at the receiver side. When the decoding order is variable, we identify which conditions the optimal power control policies should satisfy. We also describe an algorithm to determine such policies.

The remainder of the paper is organized as follows. Section II describes the system model. In Section III, effective capacity as a measure of the performance under statistical QoS constraints is briefly discussed, and the *throughput region* under QoS constraints is defined. In Section IV, under the assumption of no power control, we analyze the throughput region for both fixed and variable decoding order strategies. Section V

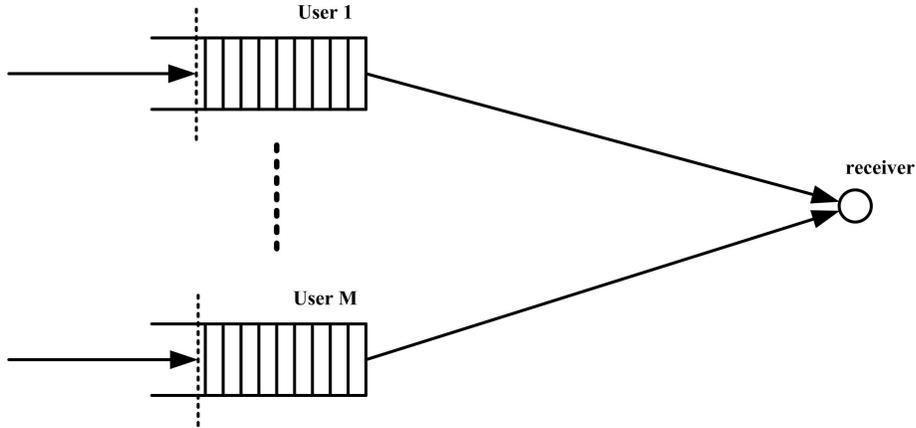

Fig. 1. The system model.

describes the optimal power control policies. Finally, Section VI concludes the paper.

## II. SYSTEM MODEL AND MAC CAPACITY REGION

As shown in Figure 1, we consider an uplink scenario where $M$ users with individual power and buffer constraints (i.e., QoS constraints) communicate with a single receiver. It is assumed that the transmitters generate data sequences which are divided into frames of duration $T$. These data frames are initially stored in the buffers before they are transmitted over the wireless channel. The discrete-time signal at the receiver in the $i^{\text{th}}$ symbol duration is given by

$$Y[i] = \sum_{j=1}^{M} h_j[i] X_j[i] + n[i], \quad i = 1, 2, \ldots \tag{1}$$

where $M$ is the number of users, $X_j[i]$ and $h_j[i]$ denote the complex-valued channel input and the fading coefficient of the $j$th user, respectively. We assume that $\{h_j[i]\}$'s are jointly stationary and ergodic discrete-time processes, and we denote the magnitude-square of the fading coefficients by $z_j[i] = |h_j[i]|^2$. Above, $n[i]$ is a zero-mean, circularly symmetric, complex Gaussian random variable with variance $\mathbb{E}\{|n[i]|^2\} = N_0$. The additive Gaussian noise samples $\{n[i]\}$ are assumed to form an independent and identically distributed (i.i.d.) sequence. Finally, $Y[i]$ denotes the received signal.

The channel input of user $j$ is subject to an average energy constraint $\mathbb{E}\{|x_j[i]|^2\} \leq \bar{P}_j/B$ for all $j$, where $B$ is the bandwidth available in the system. This formulation indicates that user $j$ is subject to an average power constraint of $\bar{P}_j$. With these definitions, the average transmitted signal to noise ratio of user $j$ is $\text{SNR}_j = \frac{\bar{P}_j}{N_0 B}$. Now, if we denote $P_j[i]$ as the instantaneous transmit power in the $i$th frame, the instantaneous transmitted SNR level becomes $\mu_j[i] = \frac{P_j[i]}{N_0 B}$. Then, the average power constraint is equivalent to the average SNR constraint $\mathbb{E}\{\mu_j[i]\} \leq \text{SNR}_j$ for user $j$.

## A. Fixed Power and Variable Rate

First, we consider the case in which the transmitters operate at fixed power and hence do not employ any power adaptation policies. The capacity region of this channel is given by [1], [4]:

$$\mathcal{R}_{\text{MAC}} = \left\{ (R_{avg,1}, \ldots, R_{avg,M}) : \mathbf{R}_{avg}(S) \leq B \mathbb{E}_{\mathbf{z}} \left\{ \log_2 \left( 1 + \sum_{j \in S} \text{SNR}_j z_j \right) \right\}, \forall S \subset \{1, \ldots, M\} \right\} \quad (2)$$

where $\text{SNR}_j = \bar{P}_j/(N_0 B)$ denotes the average transmitted signal-to-noise ratio of user $j$, $\mathbf{z} = (z_1, \cdots, z_M)$ is a random vector comprised of the magnitude-squares of the channel coefficients. As well-known, there are $M!$ vertices of the polyhedron defined in (2). The vertex $\mathbf{R}_{avg,\pi} = (R_{avg,\pi(1)}, \cdots, R_{avg,\pi(M)})$ corresponds to a permutation $\pi$, or the successive decoding order at the receiver, i.e., users are decoded in the order given by $\pi(1), \cdots, \pi(M)$. This vertex is specified by the average rates

$$R_{avg,\pi(k)} = B \mathbb{E}_{\mathbf{z}} \left\{ \log_2 \left( 1 + \frac{\text{SNR}_{\pi(k)} z_{\pi(k)}}{1 + \sum_{i=k+1}^{M} \text{SNR}_{\pi(i)} z_{\pi(i)}} \right) \right\} \text{ bits/s}, \quad k = 1, \cdots, M. \quad (3)$$

With this characterization, we see that for the given decoding order $\pi$, the maximum instantaneous service rate for user $\pi(k)$ is

$$R_{\pi(k)} = B \log_2 \left( 1 + \frac{\text{SNR}_{\pi(k)} z_{\pi(k)}}{1 + \sum_{i=k+1}^{M} \text{SNR}_{\pi(i)} z_{\pi(i)}} \right) \text{ bits/s} \quad k = 1, \cdots, M. \quad (4)$$

Finally, we note that time sharing among these $M!$ permutations of decoding orders yields any point on the boundary surface of $\mathcal{R}_{\text{MAC}}$ [18]. As also discussed in [7], it can be easily verified that varying the decoding order according to the channel states does not provide any improvement on the capacity region.

## B. Variable Power and Variable Rate

Now, we suppose that dynamic power and rate allocation is performed according to time-variations in the channels. For a given set of power allocation policies $\mathcal{U} = \{\mu_1, \cdots, \mu_M\}$, where $\mu_j \geq 0$ is the power control policy of the $j$th user, the achievable rate region is described by [4]

$$\mathcal{R}(\mathcal{U}) = \left\{ \mathbf{R}_{avg} : \mathbf{R}_{avg}(S) \leq \mathbb{E}_{\mathbf{z}} \left\{ B \log_2 \left( 1 + \sum_{j \in S} \mu_j(\mathbf{z}) z_j \right) \right\}, \forall S \subset \{1, \cdots, M\} \right\}. \tag{5}$$

For a given decoding order at the receiver, the individual average and instantaneous rates of the users can be obtained similar to (3) and (4), respectively, with SNR replaced by $\mu$. The capacity region is given by

$$\mathcal{R}_{\text{MAC}} = \bigcup_{\mathcal{U} \in \mathcal{F}} \mathcal{R}(\mathcal{U}) \tag{6}$$

where $\mathcal{F}$ is the set of all feasible power control policies that satisfy the average power constraint

$$\mathcal{F} \equiv \{\mathcal{U} : \mathbb{E}_{\mathbf{z}}\{\mu_j(\mathbf{z})\} \leq \text{SNR}_j, \mu_j \geq 0, \forall j\} \tag{7}$$

where $\text{SNR}_j = \bar{P}_j/(N_0 B)$ denotes the average transmitted signal-to-noise ratio of user $j$.

## C. TDMA

For simplicity, we assume that the time division strategy is fixed prior to transmission. Let $\delta_j$ denote the fraction of time allocated to user $j$. Note that we have $\sum_{j=1}^{M} \delta_j = 1$. In each frame, each user occupies the entire bandwidth to transmit the signal in the corresponding fraction of time. Then, the instantaneous service rate for user $j$ is given by

$$R_j(\text{SNR}_j) = B \log_2 \left( 1 + \frac{\text{SNR}_j}{\delta_j} z_j \right) \text{ bits/s}. \tag{8}$$

## III. PRELIMINARIES

### A. Effective Capacity

In [12], Wu and Negi defined the effective capacity as the maximum constant arrival rate[1] that a given service process can support in order to guarantee a statistical QoS requirement specified by the QoS exponent $\theta$. If we define $Q$ as the stationary queue length, then $\theta$ is the decay rate of the tail distribution of the queue length $Q$:

$$\lim_{q \to \infty} \frac{\log P(Q \geq q)}{q} = -\theta. \tag{9}$$

Therefore, for large $q_{\max}$, we have the following approximation for the buffer violation probability: $P(Q \geq q_{\max}) \approx e^{-\theta q_{\max}}$. Hence, while larger $\theta$ corresponds to more strict QoS constraints, smaller $\theta$ implies looser QoS guarantees. Similarly, if $D$ denotes the steady-state delay experienced in the buffer, then $P(D \geq d_{\max}) \approx e^{-\theta \xi d_{\max}}$ for large $d_{\max}$, where $\xi$ is determined by the arrival and service processes [14]. Since the average arrival rate is equal to the average departure rate when the queue is in steady-state, effective capacity can also be seen as the maximum throughput in the presence of such constraints.

The effective capacity is given by

$$C(\theta) = -\lim_{t \to \infty} \frac{1}{\theta t} \log_e \mathbb{E}\{e^{-\theta S[t]}\} \quad \text{bits/s}, \tag{10}$$

where the expectation is with respect to $S[t] = \sum_{i=1}^{t} s[i]$, which is the time-accumulated service process. $\{s[i], i = 1, 2, \ldots\}$ denote the discrete-time stationary and ergodic stochastic service process.

In this paper, in order to simplify the analysis while considering general fading distributions, we assume that the fading coefficients stay constant over the frame duration $T$ and vary independently for each frame and each user. In this scenario, $s[i] = TR[i]$, where $R[i]$ is the instantaneous service rate in the $i$th frame duration $[iT; (i+1)T)$. Then, (10) can be written as

$$C(\theta) = -\frac{1}{\theta T} \log_e \mathbb{E}_{\mathbf{z}}\{e^{-\theta T R[i]}\} \quad \text{bits/s}, \tag{11}$$

---

[1] For time-varying arrival rates, effective capacity specifies the effective bandwidth of the arrival process that can be supported by the channel.

where $R[i]$ is in general a function of the fading state $\mathbf{z}$. (11) is obtained using the fact that instantaneous rates $\{R[i]\}$ vary independently from one frame to another. It is interesting to note that as $\theta \to 0$ and hence QoS constraints relax, effective capacity approaches the ergodic rate, i.e., $C(\theta) \to \mathbb{E}_\mathbf{z}\{R[i]\}$.

Throughout the rest of the paper, we use the effective capacity normalized by bandwidth $B$, which is denoted by

$$\mathsf{C}(\theta) = \frac{C(\theta)}{B} \quad \text{bits/s/Hz}. \tag{12}$$

*B. Throughput Region*

Suppose that $\Theta = (\theta_1, \cdots, \theta_M)$ is a vector composed of the QoS constraints of $M$ users. Let $\mathsf{C}(\Theta) = (\mathsf{C}_1(\theta_1), \cdots, \mathsf{C}_M(\theta_M))$ denote the vector of the normalized effective capacities. We first have the following characterization.

*Definition 1:* The *effective throughput region* is described as

$$\mathcal{C}_{\text{MAC}}(\Theta, \text{SNR}) = \bigcup_{\substack{\mathbf{R} \\ \text{s.t. } \mathbb{E}\{\mathbf{R}\} \in \mathcal{R}_{\text{MAC}}}} \left\{ \mathsf{C}(\Theta) \geq \mathbf{0} : \mathsf{C}_j(\theta_j) \leq -\frac{1}{\theta_j TB} \log_e \mathbb{E}_\mathbf{z}\left\{ e^{-\theta TR_j} \right\} \right\} \tag{13}$$

where $\mathbf{R} = \{R_1, R_2, \cdots, R_M\}$ represents the vector composed of the instantaneous transmission (or equivalently service) rates of $M$ users. Note that the union is over the distributions of the vector $\mathbf{R}$ such that the expected value $\mathbb{E}\{\mathbf{R}\}$ lies in the MAC capacity region.

*Remark 1:* The *throughput region* given in Definition 1 represents the set of all vectors of constant arrival rates $\mathsf{C}(\theta)$ that can be supported in the fading multiple access channel in the presence QoS constraints specified by $\Theta = (\theta_1, \cdots, \theta_M)$. Since reliable communications is considered, the arrival rates are supported by instantaneous service rates whose expected values are in the MAC capacity region. For instance, in the absence of power control, the maximum instantaneous service rates for a given decoding order are given by (4).

Using the convexity of the MAC capacity region $\mathcal{R}_{\text{MAC}}$, we obtain the following preliminary result on the effective throughput region defined in (13).

*Theorem 1:* The *throughput region* $\mathcal{C}_{\text{MAC}}(\Theta, \text{SNR})$ is convex.

*Proof:* Let the vectors $\mathsf{C}(\Theta)$ and $\mathsf{C}'(\Theta)$ belong to $\mathcal{C}_{\text{MAC}}(\Theta, \textsc{snr})$. Then, there exist some rate vectors $\mathbf{R}$ and $\mathbf{R}'$ for $\mathsf{C}(\Theta)$ and $\mathsf{C}'(\Theta)$, respectively, such that $\mathbb{E}\{\mathbf{R}\}$ and $\mathbb{E}\{\mathbf{R}'\}$ are in the MAC capacity region. By a time sharing strategy, for any $\alpha \in (0,1)$, we know from the convexity of the MAC capacity region that $\mathbb{E}\{\alpha \mathbf{R} + (1-\alpha)\mathbf{R}'\} \in \mathcal{R}_{\text{MAC}}$. Now, we can write

$$\alpha \mathsf{C}(\Theta) + (1-\alpha)\mathsf{C}'(\Theta)$$

$$\leq -\frac{1}{\Theta TB} \log_e \left(\mathbb{E}\left\{e^{-\Theta T \mathbf{R}}\right\}\right)^{\alpha} \left(\mathbb{E}\left\{e^{-\Theta T \mathbf{R}'}\right\}\right)^{1-\alpha} \tag{14}$$

$$= -\frac{1}{\Theta TB} \log_e \left(\mathbb{E}\left\{\left(e^{-\Theta T \alpha \mathbf{R}}\right)^{\frac{1}{\alpha}}\right\}\right)^{\alpha} \left(\mathbb{E}\left\{\left(e^{-\Theta T (1-\alpha)\mathbf{R}'}\right)^{\frac{1}{1-\alpha}}\right\}\right)^{1-\alpha} \tag{15}$$

$$\leq -\frac{1}{\Theta TB} \log_e \mathbb{E}\left\{e^{-\Theta T(\alpha \mathbf{R} + (1-\alpha)\mathbf{R}')}\right\}. \tag{16}$$

Above, in (14) through (16), all operations, including the logarithm and exponential functions and expectations, are component-wise operations. For instance, the expression in (14) denotes a vector whose components are $\left\{\frac{1}{\theta_j TB} \log_e \left(\mathbb{E}\left\{e^{-\theta_j T R_j}\right\}\right)^{\alpha} \left(\mathbb{E}\left\{e^{-\theta T R'_j}\right\}\right)^{1-\alpha}\right\}_{j=1}^{M}$. Similarly, the inequalities in (14) and (16) are component-wise inequalities. The inequality in (14) follows from the definition in (13). Moreover, (16) follows from Hölder's inequality and leads to the conclusion that $\alpha \mathsf{C} + (1-\alpha)\mathsf{C}'$ still lies in the *throughput region*, proving the convexity result. □

We are interested in the boundary of the region $\mathcal{C}_{\text{MAC}}(\Theta, \textsc{snr})$. Now that $\mathcal{C}_{\text{MAC}}(\Theta, \textsc{snr})$ is convex, we can characterize the boundary surface by considering the following optimization problem [4]:

$$\max \lambda \cdot \mathsf{C}(\Theta) \quad \text{subject to: } \mathsf{C}(\Theta) \in \mathcal{C}_{\text{MAC}}(\Theta, \textsc{snr}). \tag{17}$$

for all priority vectors $\lambda = (\lambda_1, \cdots, \lambda_M)$ in $\mathfrak{R}_+^M$ with $\sum_{j=1}^{M} \lambda_j = 1$.

## IV. TRANSMISSIONS WITHOUT POWER CONTROL

In this section, we assume that the signals are transmitted at a constant power level in each frame and hence power adaptation with respect to the fading states is not performed. Under this assumption, we initially consider the scenario in which the receiver decodes the users in a fixed order. Subsequently, we analyze the case of variable decoding order.

## A. Fixed Decoding Order

We first assume that the receiver decodes the users in a fixed order in each frame. Hence, the decoding order does not change with respect to the realizations of the fading coefficients. If a single decoding order is used in the frame, it is obvious that only the vertices of the boundary region can be achieved. We consider a slightly more general case in which time sharing technique is employed in each frame among different decoding orders. Note that the time sharing strategy is also independent of the channel states and hence is fixed in different blocks. We denote the fraction of time allocated to decoding order $\pi_m$ as $\tau_m$. Naturally, the fractions of time satisfy $\tau_m \geq 0$ and $\sum_{m=1}^{M!} \tau_m = 1$. Varying the values of $\tau_m$ enables us to characterize the throughput region. Under these assumptions, the effective capacity for each user on the boundary surface is

$$\mathsf{C}_j(\theta_j) = -\frac{1}{\theta_j TB} \log_e \mathbb{E}_{\mathbf{z}} \left\{ e^{-\theta_j T \sum_{m=1}^{M!} \tau_m R_{\pi_m^{-1}(j)}} \right\} \tag{18}$$

where $R_{\pi_m^{-1}(j)}$ represents the maximal instantaneous service rate of user $j$ at a given decoding order $\pi_m$, which is given by

$$R_{\pi_m^{-1}(j)} = B \log_2 \left( 1 + \frac{\mathsf{SNR}_j z_j}{1 + \sum_{\pi_m^{-1}(i) > \pi_m^{-1}(j)} \mathsf{SNR}_i z_i} \right) \tag{19}$$

where $\pi_m^{-1}$ is the inverse trace function of $\pi_m$.

*Remark 2:* Note that $R_{\pi_m^{-1}(j)}$ is the maximum instantaneous service rate achieved with superposition coding and a particular decoding order. Hence, the corresponding effective capacities characterize the throughput achieved with this strategy in the presence of QoS constraints.

*Remark 3:* Throughout the rest of the paper, we generally specify the effective capacity values on the boundary surface for simplicity and brevity. Effective capacity regions can immediately be specified using these boundary points. For instance, the effective capacity (or equivalently throughput) region for superposition coding and fixed decoding order is

$$\bigcup_{\{\tau_m\}} \left\{ \mathsf{C}(\Theta) \geq \mathbf{0} : \mathsf{C}_j(\theta_j) \leq -\frac{1}{\theta_j TB} \log_e \mathbb{E}_{\mathbf{z}} \left\{ e^{-\theta_j T \sum_{m=1}^{M!} \tau_m R_{\pi_m^{-1}(j)}} \right\} \right\} \tag{20}$$

where the union is over different time allocation strategies.

Next, for comparison, we consider the TDMA case in which we also have similar time allocation strategies but only one user transmits in its specific fraction of time. We first have the following definition.

*Definition 2:* The *throughput region* for TDMA can be seen as the achievable vectors of arrival rates with each component bounded by the effective capacity obtained when the instantaneous service rate is given by (8). More specifically, the maximum effective capacity for user $j$ is

$$\mathsf{C}_j^{\mathrm{TD}}(\theta_j) = -\frac{1}{\theta_j TB} \log_e \mathbb{E}\left\{ e^{-\delta_j \theta_j TB \log_2\left(1+\frac{\mathrm{SNR}_j}{\delta_j}z_j\right)} \right\} \quad (21)$$

where $\delta_j$ is the fraction of time allocated to user $j$, and $0 \leq \delta_j \leq 1$.

An immediate result can be obtained as follows:

*Theorem 2:* The *throughput region* for TDMA is convex.

*Proof:* Note that the points on the boundary surface is given in (21). Consider the function $f(\delta) = -\delta \theta TB \log_2\left(1 + \frac{\mathrm{SNR}}{\delta}z\right)$. It can be easily verified that $f(\delta)$ is a convex function in $\delta$. Then, $e^{f(\delta)}$ is a log-convex function. Since weighted non-negative sum preserves the log-convexity [19, Section 3.5], we know that $\mathbb{E}_z\{e^{f(\delta)}\}$ is log-convex. Then $-\frac{1}{\theta TB}\log_e \mathbb{E}\{e^{-\delta \theta TB \log_2\left(1+\frac{\mathrm{SNR}}{\delta}z\right)}\}$ is a concave function in $\delta$. Hence, we immediately see that the *throughput region* for TDMA is convex. $\square$

The optimal time allocation policy that maximizes the weighted sum can be obtained through the optimization problem

$$\max_{\{\delta_j\}} \sum_{j=1}^{M} -\frac{\lambda_j}{\theta_j TB} \log_e \mathbb{E}\left\{ e^{-\delta_j \theta_j TB \log_2\left(1+\frac{\mathrm{SNR}_j}{\delta_j}z_j\right)} \right\}, \text{ s.t. } \sum_{j=1}^{M} \delta_j = 1, \delta_j \geq 0. \quad (22)$$

The objective function in the above problem is concave, and we can use the Lagrangian maximization approach. Taking the derivative of the Lagrangian function with respect to $\delta_j$, we obtain the following optimality condition for each user:

$$\frac{\partial \mathcal{J}}{\partial \delta_j} = \lambda_j \frac{\mathbb{E}\left\{ e^{-\delta_j \theta_j TB \log_2\left(1+\frac{\mathrm{SNR}_j}{\delta_j}z_j\right)} \left(\log_2(1+\frac{\mathrm{SNR}_j}{\delta_j}z_j) - \frac{\frac{\mathrm{SNR}_j}{\delta_j}z_j}{1+\frac{\mathrm{SNR}_j}{\delta_j}z_j}\log_2 e\right) \right\}}{\mathbb{E}\left\{ e^{-\delta_j \theta_j TB \log_2\left(1+\frac{\mathrm{SNR}_j}{\delta_j}z_j\right)} \right\}} - \kappa = 0 \quad (23)$$

where $\kappa$ is the Lagrange multiplier whose value is chosen to satisfy the constraint $\sum_{j=1}^{M} \delta_j = 1$. If the optimal value of $\delta_j$ turns out to be negative, then the optimal value of $\delta_j$ should be 0. When $\lambda_1 = \lambda_2 = \cdots = \lambda_M$, the obtained values of $\{\delta_j\}$ are the ones that achieve the maximal sum-rate throughput, i.e., the sum of the effective capacities of the users. Although obtaining closed-form solutions is unlikely, the maximization problem in (22) can be easily solved numerically using convex optimization tools. Numerical results are provided in Section IV-C.

*B. Variable Decoding Order*

We now study the case in which the receiver varies the decoding order with respect to the fading states $\mathbf{z} = (z_1, \ldots, z_M)$. More specifically, we assume that the vector space $\mathfrak{R}_+^M$ of the possible values for $\mathbf{z}$ is partitioned into $M!$ disjoint regions $\{\mathcal{Z}_m\}_{m=1}^{M!}$ with respect to decoding orders $\{\pi_m\}_{m=1}^{M!}$. Hence, each region corresponds to a unique decoding order. For instance, when $\mathbf{z} \in \mathcal{Z}_1$, the receiver decodes the information in the order $\pi_1$. Now, for a given partition $\{\mathcal{Z}_m\}_{m=1}^{M!}$, the maximum effective capacity that can be achieved by the $j$th user is

$$\mathsf{C}_j(\theta_j) = -\frac{1}{\theta_j T B} \log_e \mathbb{E}_{\mathbf{z}} \left\{ e^{-\theta_j T R_j} \right\} \tag{24}$$

$$= -\frac{1}{\theta_j T B} \log_e \left( \sum_{m=1}^{M!} \int_{\mathbf{z} \in \mathcal{Z}_m} e^{-\theta_j T R_{\pi_m^{-1}(j)}} p_{\mathbf{z}}(\mathbf{z}) d\mathbf{z} \right) \text{ for } j = 1, 2, \ldots, M \tag{25}$$

where $p_{\mathbf{z}}$ is the distribution function of $\mathbf{z}$ and $R_{\pi_m^{-1}(j)}$ is given in (19). Akin to the optimization in (17), the optimal partition $\{\mathcal{Z}_m\}_{m=1}^{M!}$ that maximizes the weighted sum of the effective capacities can be identified by solving the following optimization problem:

$$\max_{\{\mathcal{Z}_m\}} \boldsymbol{\lambda} \cdot \mathsf{C}(\boldsymbol{\Theta}) = \max_{\{\mathcal{Z}_m\}} \sum_{j=1}^{M} \lambda_j \mathsf{C}_j(\theta_j) \tag{26}$$

$$= \max_{\{\mathcal{Z}_m\}} \sum_{j=1}^{M} -\frac{\lambda_j}{\theta_j T B} \log_e \left( \sum_{m=1}^{M!} \int_{\mathbf{z} \in \mathcal{Z}_m} e^{-\theta_j T R_{\pi_m^{-1}(j)}} p_{\mathbf{z}}(\mathbf{z}) d\mathbf{z} \right). \tag{27}$$

Note that the optimal partition depends on the weight vector $\boldsymbol{\lambda} = (\lambda_1, \ldots, \lambda_M)$. By solving a sequence of optimization problems for different values of $\boldsymbol{\lambda}$, we can trace the boundary of the effective throughput region.

Considering the expression for effective capacity and the optimization problem in (27), we note that finding closed-form analytical expressions for the optimal partitions of the channel state space seems intractable for a general scenario. With this in mind, we consider a simplified case in which all users have the same QoS constraint described by $\theta$. This case arises, for instance, if users do not have priorities over others in terms of buffer limitations or delay constraints.

*1) Two-user MAC:* First, we consider the two-user MAC case and suppose that the two users have the same QoS exponent $\theta$. Similar to the discussion in [17], finding an optimal decoding order function can be reduced to finding a function $z_2 = g(z_1)$ in the state space such that users are decoded in the order (1,2) if $z_2 < g(z_1)$ and users are decoded in the order (2,1) if $z_2 > g(z_1)$. Hence, the function $g$ partitions the space of the possible values of $\mathbf{z} = (z_1, z_2)$. With this, the optimization problem in (26) becomes

$$\max_{g} \lambda_1 \mathsf{C}_1(\theta, g(z_1)) + (1 - \lambda_1) \mathsf{C}_2(\theta, g(z_1)) \tag{28}$$

where $\mathsf{C}_1(\theta, g(z_1))$ and $\mathsf{C}_2(\theta, g(z_1))$ are expressed as

$$\mathsf{C}_1(\theta, g(z_1)) = -\frac{1}{\theta TB} \log_e \left( \int_0^\infty \int_{g(z_1)}^\infty e^{-\theta TB \log_2(1 + \mathrm{SNR}_1 z_1)} p_\mathbf{z}(z_1, z_2) dz_2 dz_1 \right.$$
$$\left. + \int_0^\infty \int_0^{g(z_1)} e^{-\theta TB \log_2 \left(1 + \frac{\mathrm{SNR}_1 z_1}{1 + \mathrm{SNR}_2 z_2}\right)} p_\mathbf{z}(z_1, z_2) dz_2 dz_1 \right) \tag{29}$$

$$\mathsf{C}_2(\theta, g(z_1)) = -\frac{1}{\theta TB} \log_e \left( \int_0^\infty \int_0^{g(z_1)} e^{-\theta TB \log_2(1 + \mathrm{SNR}_2 z_2)} p_\mathbf{z}(z_1, z_2) dz_2 dz_1 \right.$$
$$\left. + \int_0^\infty \int_{g(z_1)}^\infty e^{-\theta TB \log_2 \left(1 + \frac{\mathrm{SNR}_2 z_2}{1 + \mathrm{SNR}_1 z_1}\right)} p_\mathbf{z}(z_1, z_2) dz_2 dz_1 \right). \tag{30}$$

Note that the maximization in (28) is over the choice of the function $g(z_1)$. Implicitly, $g(z_1)$ should always be larger than zero as implicitly implied in (29) and (30). In cases in which this condition is not satisfied, we need to find a function $z_1 = f(z_2)$ instead, as will be specified below.

*Theorem 3:* The optimal decoding order as a function of the fading state $\mathbf{z} = (z_1, z_2)$ for a specific common QoS constraint $\theta$ in the two-user case is characterized by the following functions:

$$g(z_1) = \frac{(1 + \text{SNR}_1 z_1) K^{\frac{1}{\beta}} - 1}{\text{SNR}_2}, \quad \text{if } K \in [1, \infty) \text{ and} \tag{31}$$

$$f(z_2) = \frac{(1 + \text{SNR}_2 z_2) K^{-\frac{1}{\beta}} - 1}{\text{SNR}_1}, \quad \text{if } K \in [0, 1) \tag{32}$$

where $\beta = \frac{\theta TB}{\log_e 2}$ and $K \in [0, \infty)$ is a constant that depends on the weight $\lambda_1$ in (28) and the values of the double integrals in (29) and (30). Note that the function used to partition the state space is either $g$ or $f$ depending on the value of $K$.

*Proof:* Suppose that the optimal decoding order is specified by the function $z_2 = g(z_1)$. We define

$$\mathcal{J}(\hat{g}(z_1)) = \lambda_1 \mathsf{C}_1(\theta, \hat{g}(z_1)) + (1 - \lambda_1) \mathsf{C}_2(\theta, \hat{g}(z_1)) \tag{33}$$

where $\hat{g}(z_1) = g(z_1) + s\eta(z_1)$. $g(z_1)$ is the optimal function, $s$ is any constant, and $\eta(z_1)$ represents arbitrary perturbation. A necessary condition that needs to be satisfied is [20]

$$\left. \frac{d}{ds} (\mathcal{J}(\hat{g}(z_1))) \right|_{s=0} = 0. \tag{34}$$

We define the following:

$$\phi_1 = \int_0^\infty \int_{g(z_1)}^\infty e^{-\theta TB \log_2(1 + \text{SNR}_1 z_1)} p_\mathbf{z}(z_1, z_2) dz_2 dz_1 + \int_0^\infty \int_0^{g(z_1)} e^{-\theta TB \log_2 \left(1 + \frac{\text{SNR}_1 z_1}{1 + \text{SNR}_2 z_2}\right)} p_\mathbf{z}(z_1, z_2) dz_2 dz_1 \tag{35}$$

$$\phi_2 = \int_0^\infty \int_0^{g(z_1)} e^{-\theta TB \log_2(1 + \text{SNR}_2 z_2)} p_\mathbf{z}(z_1, z_2) dz_2 dz_1 + \int_0^\infty \int_{g(z_1)}^\infty e^{-\theta TB \log_2 \left(1 + \frac{\text{SNR}_2 z_2}{1 + \text{SNR}_1 z_1}\right)} p_\mathbf{z}(z_1, z_2) dz_2 dz_1 \tag{36}$$

By noting that $\frac{d\hat{g}(z_1)}{ds} = \eta(z_1)$, and from (34)–(36), we can derive

$$\int_0^\infty \left( -\frac{\lambda_1}{\theta TB\phi_1} \left( \left(1 + \frac{\text{SNR}_1 z_1}{1 + \text{SNR}_2 g(z_1)}\right)^{-\beta} - (1 + \text{SNR}_1 z_1)^{-\beta} \right) \right.$$
$$\left. -\frac{1-\lambda_1}{\theta TB\phi_2} \left( (1 + \text{SNR}_2 g(z_1))^{-\beta} - \left(1 + \frac{\text{SNR}_2 g(z_1)}{1 + \text{SNR}_1 z_1}\right)^{-\beta} \right) \right)$$
$$\cdot p_{\mathbf{z}}(z_1, g(z_1))\eta(z_1) dz_1 = 0 \quad (37)$$

Since the above equation holds for any $\eta(z_1)$, it follows that

$$-\frac{\lambda_1}{\theta TB\phi_1} \left( \left(1 + \frac{\text{SNR}_1 z_1}{1 + \text{SNR}_2 g(z_1)}\right)^{-\beta} - (1 + \text{SNR}_1 z_1)^{-\beta} \right)$$
$$-\frac{1-\lambda_1}{\theta TB\phi_2} \left( (1 + \text{SNR}_2 g(z_1))^{-\beta} - \left(1 + \frac{\text{SNR}_2 g(z_1)}{1 + \text{SNR}_1 z_1}\right)^{-\beta} \right) = 0 \quad (38)$$

which after rearranging and defining $K$ as follows yields

$$\frac{\left(1 + \frac{\text{SNR}_1 z_1}{1 + \text{SNR}_2 g(z_1)}\right)^{-\beta} - (1 + \text{SNR}_1 z_1)^{-\beta}}{\left(1 + \frac{\text{SNR}_2 g(z_1)}{1 + \text{SNR}_1 z_1}\right)^{-\beta} - (1 + \text{SNR}_2 g(z_1))^{-\beta}} = \frac{(1-\lambda_1)\phi_1}{\lambda_1 \phi_2} = K. \quad (39)$$

Obviously, $K \geq 0$. Notice that after a simple computation, (39) becomes

$$\left( \frac{1 + \text{SNR}_1 z_1}{1 + \text{SNR}_2 g(z_1)} \right)^{-\beta} = K \quad (40)$$

which leads to (31) after rearranging. Note here that if $K < 1$, $g(z_1) < 0$ for $z_1 < \frac{K^{-\frac{1}{\beta}}-1}{\text{SNR}_1}$. Then, the expressions in (29) and (30) are not well-defined. In this case, we denote the optimal function as $z_1 = f(z_2)$ instead. Following a similar approach as shown in (29) through (40) yields (32). $\square$

*Remark 4:* Above, we have assumed that the users are decoded in the order $(1, 2)$ when $z_2 < g(z_1)$ (or $z_1 > f(z_2)$ if $K < 1$) and decoded in the order $(2, 1)$ when $z_2 > g(z_1)$ (or $z_1 < f(z_2)$ if $K < 1$). It is interesting note that if we switch the decoding orders in the regions (i.e., if users are decoded in the order $(1, 2)$ when $z_2 > g(z_1)$), exactly the same partition functions as in (31) and (32) are obtained due to the symmetric nature of the problem. Hence, the structure of the optimal functions that partition the space of channel states $(z_1, z_2)$ into two non-overlapping regions do not depend on which decoding order is used in which region.

*Remark 5:* Although the partition does not depend on the choice of the decoding orders in different regions, the performance definitely does. Our numerical computations show that the order selected originally at the beginning of our discussion (i.e., using the decoding order (1,2) when $z_2 < g(z_1)$ or $z_1 > f(z_2)$) provides a larger throughput region than otherwise. This observation leads to an interesting conclusion. Note that partition functions $g(z_1)$ in (31) and $f(z_2)$ in (32) are linear functions of $z_1$ and $z_2$, respectively. When $K \geq 1$ and

$$z_2 < g(z_1) = \frac{(1 + \text{SNR}_1 z_1) K^{\frac{1}{\beta}} - 1}{\text{SNR}_2}, \tag{41}$$

user 1 is decoded first and user 2 is decoded last. Hence, for instance, when $z_1$ is much larger than $z_2$ and user 1 is enjoying much better channel conditions, user 1 is decoded first in the presence of interference caused by user 2's received signal. User 2, who has less favorable conditions, is decoded subsequently without experiencing any interference. Note that such an operation is the opposite of an opportunistic behavior and leads to a more fair treatment of users. This is rather insightful since the users are assumed to operate under similar QoS limitations (i.e., they have the same QoS exponent $\theta$). Note that if the decoding orders are switched, users having favorable channel conditions will be decoded last and hence experience no interference. In such a case, there is a bias towards users with better channel conditions, which leads to inefficient performance when both users operate under similar buffer constraints.

Our observations above have led us to propose the following suboptimal decoding order strategy for a scenario with more than 2 users.

*2) Suboptimal Decoding Order:* In this section, we consider an arbitrary number of users. When all users have the same QoS constraint specified by $\theta$, we propose a suboptimal decoding order given by

$$\frac{\lambda_{\pi(1)}}{z_{\pi(1)}} \leq \frac{\lambda_{\pi(2)}}{z_{\pi(2)}} \cdots \leq \frac{\lambda_{\pi(M)}}{z_{\pi(M)}}, \tag{42}$$

due to the observation that the user with the largest weight $\lambda$ should be decoded last, and the fact that the higher the value of $z$, the less power is needed to achieve a specific effective capacity. Considering a two-user example, we, with this choice of the decoding order, can express the points on the boundary surface as

$$\mathsf{C}_1(\theta) = -\frac{1}{\theta TB} \log_e \left( \int_0^\infty \int_{\frac{\lambda_2 z_1}{\lambda_1}}^\infty e^{-\theta TB \log_2(1+\mathsf{SNR}_1 z_1)} p_\mathbf{z}(z_1, z_2) dz_2 dz_1 \right.$$
$$\left. + \int_0^\infty \int_0^{\frac{\lambda_2 z_1}{\lambda_1}} e^{-\theta TB \log_2\left(1+\frac{\mathsf{SNR}_1 z_1}{1+\mathsf{SNR}_2 z_2}\right)} p_\mathbf{z}(z_1, z_2) dz_2 dz_1 \right) \quad (43)$$

$$\mathsf{C}_2(\theta) = -\frac{1}{\theta TB} \log_e \left( \int_0^\infty \int_0^{\frac{\lambda_2 z_1}{\lambda_1}} e^{-\theta TB \log_2(1+\mathsf{SNR}_2 z_2)} p_\mathbf{z}(z_1, z_2) dz_2 dz_1 \right.$$
$$\left. + \int_0^\infty \int_{\frac{\lambda_2 z_1}{\lambda_1}}^\infty e^{-\theta TB \log_2\left(1+\frac{\mathsf{SNR}_2 z_2}{1+\mathsf{SNR}_1 z_1}\right)} p_\mathbf{z}(z_1, z_2) dz_2 dz_1 \right). \quad (44)$$

*C. Numerical Results*

We have performed numerical analysis for independent Rayleigh fading channels with $\mathbb{E}\{\mathbf{z}\} = \mathbf{1}$. In Fig. 2, the throughput region of a two-user MAC is plotted for superposition strategies with different decoding ordering methods at the receiver, and also for TDMA. In the figure, the solid and dotted curves provide the throughput regions achieved by employing optimal and suboptimal variable decoding orders, respectively, at the receiver. Note that in the optimal strategy described by the results of Theorem 3, the receiver chooses the decoding order according to the channel states such that the weighted sum of effective capacities, i.e., summation of $\log$-moment generating functions, is maximized. We see that the suboptimal strategy described in Section IV-B.2 can achieve almost the same rate region as the optimal strategy, indicating the efficiency of this approach. In the same figure, dot-dashed curve provides the throughput region achieved by employing a fixed decoding order for all channel states. Here, we observe that the strategy of using a fixed decoding order at the receiver is strictly suboptimal even when the users are operating under similar buffer constraints, and varying the decoding order with the respect to the channel gains can significantly increase the achievable region. Finally, the throughput region of TDMA is given by the dashed curve. We immediately note that TDMA can achieve some points outside of the throughput region attained with fixed decoding order at the receiver side. These numerical results show that markedly different strategies may need to be employed when systems are operating under buffer constraints. In the absence of such constraints, the performance is captured by the ergodic capacity region which cannot be improved by varying the decoding order with respect to the channel states [7]. Hence, using a fixed decoding order at the receiver is an optimal strategy when there are no QoS constraints. Moreover, TDMA is always suboptimal with respect to the superposition schemes regardless of the decoding-order strategy [8].

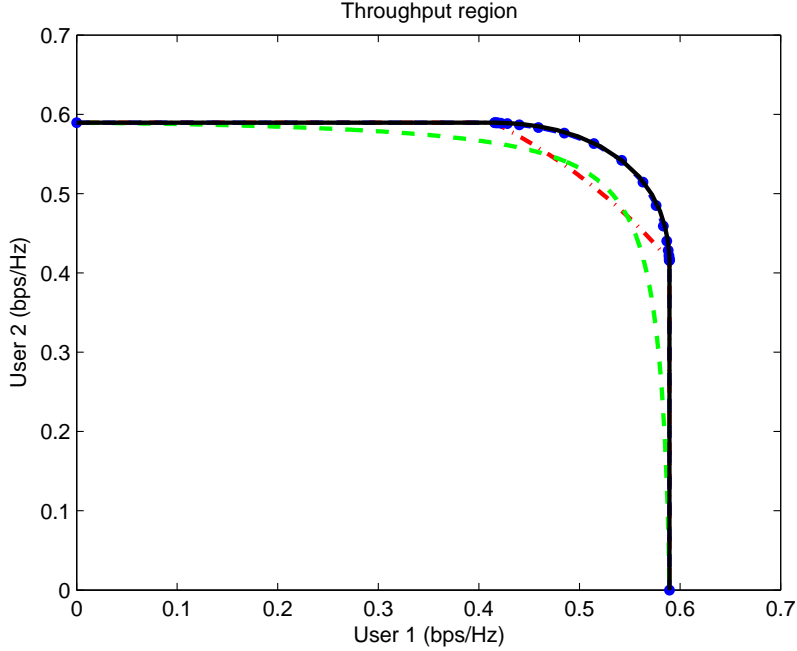

Fig. 2. The throughput region of two-user MAC case. $\text{SNR}_1 = \text{SNR}_2 = 0$ dB. $\theta_1 = \theta_2 = 0.01$. The solid, dotted, dot-dashed, and dashed lines represent the regions achieved with optimal variable decoding order, suboptimal variable decoding order, fixed decoding with time sharing, and the TDMA respectively.

In Fig. 3, sum-rate throughput, i.e. the sum of the effective capacities, is plotted as a function of the QoS exponent $\theta$. Here, we note that as $\theta$ increases, the curves of different strategies converge. In particular, TDMA performance approaches that of the superposition coding with variable decoding. Hence, orthogonal transmission strategies start being efficient in terms of attaining the sum rate under stringent buffer constraints. Note that the sum-rate throughput generally decreases with increasing $\theta$, and we conclude from the figure that this diminished throughput can be captured by having each user concentrate its power in a certain fraction of time in the TDMA scheme. We also see that for approximately $\theta > 0.006$, TDMA starts outperforming superposition transmission when a fixed decoding order is employed at the receiver. Such an observation is also noted in the discussion of Fig. 2. In contrast, we observe that as $\theta$ approaches 0 and hence the QoS constraints relax, TDMA is the strategy with the worst performance. Note that when the performance metric is the ergodic capacity and hence no queueing constraints are considered, this suboptimality of TDMA with respect to superposition strategies is well-known (see e.g., [8]).

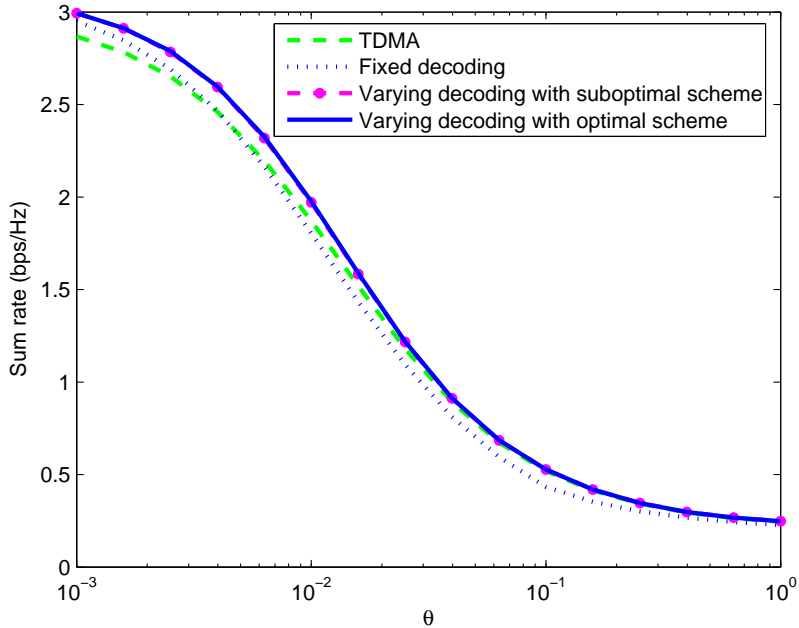

Fig. 3. The sum-rate throughput as a function of $\theta$. $\text{SNR}_1 = 10$ dB; $\text{SNR}_2 = 0$ dB.

We are also interested in the values of parameter $K$ that appear in the functions in Theorem 3. In Fig. 4, we plot $K$ as a function of $\frac{\lambda_1}{\lambda_2} = \frac{\lambda_1}{1-\lambda_1}$. It is interesting to note that $\log_e K$ seems to be linear with respect to $\log_e\left(\frac{\lambda_1}{1-\lambda_1}\right)$.

## V. TRANSMISSIONS WITH POWER CONTROL

In this section, we analyze the case in which the transmitter employs power control policies in the transmission. Similarly as before, we initially investigate the scenario in which the decoding order is fixed for all channel states. Subsequently, we study variable decoding order schemes. Note that varying the decoding order with respect to the channel states, according to the analysis in Section IV, has the potential to significantly affect the achievable rates.

Fig. 4. $K$ vs. $\frac{\lambda_1}{\lambda_2}$. $\text{SNR}_1 = 10dB$. $\text{SNR}_2 = 0$ dB. $\theta_1 = \theta_2 = 0.01$.

*A. Power Control Policy for Fixed Decoding Order in All Channel States*

Here, we characterize the optimal power allocation policies when the decoding order is fixed for all channel states. Due to the convexity of $\mathcal{C}_{\text{MAC}}$, there exist Lagrange multipliers $\kappa = (\kappa_1, \ldots, \kappa_M) \in \mathfrak{R}_+^M$ such that $\mathsf{C}^*(\Theta)$ on the boundary surface can be obtained by solving the optimization problem

$$\max_{\mu} \lambda \cdot \mathsf{C}(\Theta) - \kappa \cdot \mathbb{E}\{\mu\} \tag{45}$$

where $\mu = (\mu_1, \ldots, \mu_M)$ represents the collection of the power control policies of all users, $\lambda = (\lambda_1, \ldots, \lambda_M)$ is the weight vector, and $\mathsf{C}(\Theta) = (\mathsf{C}_1(\theta_1), \ldots, \mathsf{C}_M(\theta_M))$ is the vector of maximum effective capacities of the users for given decoding order and power allocation policies. Note that $\mu_j = \frac{P_j}{N_0 B}$ (defined in Section II as the instantaneous transmitted SNR level) describes the power control policy of the $j$th user . For a given permutation $\pi$ and set of power allocations $\mu$, $\mathsf{C}_j(\theta_j)$ is given by

$$\mathsf{C}_j(\theta_j) = -\frac{1}{\theta_j TB} \log_e \mathbb{E}\left\{e^{-\theta_j TB \log_2\left(1+\frac{\mu_j z_j}{1+\sum_{\pi^{-1}(i)>\pi^{-1}(j)} \mu_i z_i}\right)}\right\}. \qquad (46)$$

Now, the optimization problem (45) can rewritten as

$$\max_{\mu} \sum_{j=1}^{M} -\lambda_j \frac{1}{\theta_j TB} \log_e \mathbb{E}\left\{e^{-\theta_j TB \log_2\left(1+\frac{\mu_j z_j}{1+\sum_{\pi^{-1}(i)>\pi^{-1}(j)} \mu_i z_i}\right)}\right\} - \sum_{j=1}^{M} \kappa_j \mathbb{E}\{\mu_j\}. \qquad (47)$$

The following result identifies the optimal power adaptation policies that solve the above optimization problem.

*Theorem 4:* Assume that the receiver, for all channel states, decodes the users in a fixed order specified by the permutation $\pi$. Then, the optimal power allocation allocation policies that solve the optimization problem in (47) are given by

$$\mu_j = \left(\frac{\left(1+\sum_{\pi^{-1}(i)>\pi^{-1}(j)} \mu_i z_i\right)^{\frac{\beta_j}{\beta_j+1}}}{\alpha_j^{\frac{1}{\beta_j+1}} z_j^{\frac{\beta_j}{\beta_j+1}}} - \frac{1+\sum_{\pi^{-1}(i)>\pi^{-1}(j)} \mu_i z_i}{z_j}\right)^+ \quad \text{for } j = 1, 2, \ldots, M \qquad (48)$$

where $\beta_j = \frac{\theta_j TB}{\log_e 2}$ is the normalized QoS exponent, $(x)^+ = \max\{x, 0\}$, and $(\alpha_1, \cdots, \alpha_M)$ are constants that are introduced to satisfy the average power constraints.

*Proof:* Note that with a fixed decoding order, the user $\pi(M)$ sees no interference from the other users, and hence the derivative of (47) with respect to $\mu_{\pi(M)}$ will only be related to the effective capacity formulation of user $\pi(M)$. Therefore, we can solve an equivalent problem by maximizing $\mathsf{C}_{\pi(M)}$ instead. After we derive $\mu_{\pi(M)}$, the derivative of (47) with respect to $\mu_{\pi(M-1)}$ will only be related to the effective capacity formulation of user $\pi(M-1)$. By repeated application of this procedure, for given $\lambda$, (47) can be further decomposed into the following $M$ sequential optimization problems

$$\max_{\mu_j} -\lambda_j \frac{1}{\theta_j TB} \log_e \mathbb{E}\left\{e^{-\theta_j TB \log_2\left(1+\frac{\mu_j z_j}{1+\sum_{\pi^{-1}(i)>\pi^{-1}(j)} \mu_i z_i}\right)}\right\} - \kappa_j \mathbb{E}\{\mu_j\} \quad j \in \{1, \cdots, M\} \qquad (49)$$

in the inverse order of $\pi$. Similarly as in [13], due to the monotonicity of the logarithm, solving the above $M$ optimizations is the same as solving

$$\min_{\mu_j} \mathbb{E}\left\{ e^{-\theta_j TB \log_2 \left(1+\frac{\mu_j z_j}{1+\sum_{\pi^{-1}(i)>\pi^{-1}(j)} \mu_i z_i}\right)} \right\} + \kappa_j \mathbb{E}\{\mu_j\} \quad j \in \{1,\cdots,M\}. \tag{50}$$

Differentiating the above Lagrangian with respect to $\mu_j$ and setting the derivative to zero yield the intended result in (48). $\square$

*Remark 6:* Exploiting the result in (48), we can find that instead of adapting the power according to only its channel state as in [13] where a single-user scenario is studied, the user adapts the power with respect to its channel state normalized by the observed interference and the noise.

*Remark 7:* To give an explicit idea of the power control policy, we consider a two-user example in which the decoding order is $(2,1)$. For this case, we can easily find that

$$\mu_1 = \begin{cases} \dfrac{1}{\alpha_1^{\frac{1}{\beta_1+1}} z_1^{\frac{\beta_1}{\beta_1+1}}} - \dfrac{1}{z_1} & z_1 > \alpha_1, \\ 0 & \text{otherwise} \end{cases}, \tag{51}$$

and

$$\mu_2 = \begin{cases} \dfrac{1}{\alpha_2^{\frac{1}{\beta_2+1}} z_2^{\frac{\beta_2}{\beta_2+1}}} - \dfrac{1}{z_2} & z_1 \leq \alpha_1 \text{ and } z_2 > \alpha_2, \\ \dfrac{\left(\frac{z_1}{\alpha_1}\right)^{\frac{\beta_2}{(\beta_1+1)(\beta_2+1)}}}{\alpha_2^{\frac{1}{\beta_2+1}} z_2^{\frac{\beta_2}{\beta_2+1}}} - \dfrac{\left(\frac{z_1}{\alpha_1}\right)^{\frac{1}{\beta_1+1}}}{z_2} & z_1 > \alpha_1 \text{ and } \frac{z_2}{\alpha_2} > \left(\frac{z_1}{\alpha_1}\right)^{\frac{1}{\beta_1+1}}, \\ 0 & \text{otherwise} \end{cases}, \tag{52}$$

where $\alpha_1$ and $\alpha_2$ are chosen to satisfy the average power constraints of the two users.

*B. Power Control Policy for Variable Decoding Order*

In this section, we study the optimal power allocation policy when the receiver varies the decoding order with respect to the channel fading states. We mainly concentrate on the two-user scenario. The key idea we introduce here is to consider the power allocation policy of each user $j$ for each region $\mathcal{Z}_m$ (in which decoding is performed according to permutation $\pi_m$) while requiring the average power constraint to be satisfied by the joint power over all regions $\{\mathcal{Z}_m\}$.

For the two-user case, due to the convexity of the throughput region, there exist Lagrange multipliers $\kappa = (\kappa_1, \kappa_2) \in \mathfrak{R}_+^2$ such that $\mathsf{C}^*(\Theta)$ on the boundary surface can be obtained by solving the optimization problem

$$\max_{\mu} \lambda_1 \mathsf{C}_1(\mu, \mathcal{Z}) + \lambda_2 \mathsf{C}_2(\mu, \mathcal{Z}) - \kappa_1 \mathbb{E}\{\mu_1\} - \kappa_2 \mathbb{E}\{\mu_2\} \tag{53}$$

where $\mu = (\mu_1, \mu_2)$ are the power control policies, $(\lambda_1, \lambda_2)$ are the weights in the weighted sum, and $\mathcal{Z} = (\mathcal{Z}_1, \mathcal{Z}_2)$ denotes a particular partition of the space of the positive values of $\mathbf{z} = (z_1, z_2)$ [2]. Hence, power control policies that solve (53) are the optimal ones for a given partition. In the following, since we assume $\mathcal{Z}$ is given, the notation $\mathsf{C}_j(\mu, \mathcal{Z})$ is replaced by $\mathsf{C}_j(\mu)$ for brevity.

Recalling the discussion in Section IV-B, we can express the effective capacities of the two users as in (29) and (30) by only replacing $\mathsf{SNR}_j$ with $\mu_j(\mathbf{z})$ in these expressions. The Lagrangian (which is the objective function in (53)) can now be expressed as

$$\mathcal{J} = -\frac{\lambda_1}{\beta_1 \log_e 2} \log_e \left( \int_{\mathbf{z} \in \mathcal{Z}_1} \left(1 + \frac{\mu_1 z_1}{1 + \mu_2 z_2}\right)^{-\beta_1} p_{\mathbf{z}}(z_1, z_2) dz_1 dz_2 + \int_{\mathbf{z} \in \mathcal{Z}_2} (1 + \mu_1 z_1)^{-\beta_1} p_{\mathbf{z}}(z_1, z_2) dz_1 dz_2 \right)$$
$$- \frac{\lambda_2}{\beta_2 \log_e 2} \log_e \left( \int_{\mathbf{z} \in \mathcal{Z}_2} \left(1 + \frac{\mu_2 z_2}{1 + \mu_1 z_1}\right)^{-\beta_2} p_{\mathbf{z}}(z_1, z_2) dz_1 dz_2 + \int_{\mathbf{z} \in \mathcal{Z}_1} (1 + \mu_2 z_2)^{-\beta_2} p_{\mathbf{z}}(z_1, z_2) dz_1 dz_2 \right)$$
$$- \kappa_1 (\mathbb{E}_{\mathbf{z} \in \mathcal{Z}_1}\{\mu_1\} + \mathbb{E}_{\mathbf{z} \in \mathcal{Z}_2}\{\mu_1\}) - \kappa_2 (\mathbb{E}_{\mathbf{z} \in \mathcal{Z}_1}\{\mu_2\} + \mathbb{E}_{\mathbf{z} \in \mathcal{Z}_2}\{\mu_2\}). \tag{54}$$

Above, the expressions in regions $\mathcal{Z}_1$ and $\mathcal{Z}_2$ are written separately due to the reason that possibly different power allocation strategies are employed in different regions. We define

$$\phi_1 = \int_{\mathbf{z} \in \mathcal{Z}_1} \left(1 + \frac{\mu_1 z_1}{1 + \mu_2 z_2}\right)^{-\beta_1} p_{\mathbf{z}}(z_1, z_2) dz_1 dz_2 + \int_{\mathbf{z} \in \mathcal{Z}_2} (1 + \mu_1 z_1)^{-\beta_1} p_{\mathbf{z}}(z_1, z_2) dz_1 dz_2, \tag{55}$$

---

[2]Similarly as discussed in Section IV-B, different decoding orders are employed in $\mathcal{Z}_1$ and $\mathcal{Z}_2$.

and

$$\phi_2 = \int_{\mathbf{z} \in \mathcal{Z}_2} \left(1 + \frac{\mu_2 z_2}{1 + \mu_1 z_1}\right)^{-\beta_2} p_{\mathbf{z}}(z_1, z_2) dz_1 dz_2 + \int_{\mathbf{z} \in \mathcal{Z}_1} (1 + \mu_2 z_2)^{-\beta_2} p_{\mathbf{z}}(z_1, z_2) dz_1 dz_2. \quad (56)$$

Note that the values of these functions are obtained for given power control policies $\mu = (\mu_1, \mu_2)$ and given partition $\mathcal{Z} = (\mathcal{Z}_1, \mathcal{Z}_2)$.

Now, we consider the power control policy of each user in each decoding order region $\mathcal{Z}_i$, $i = 1, 2$. By differentiating the Lagrangian, we can find the following optimality conditions:

1) $\dfrac{\lambda_1}{\phi_1 \log_e 2}(1 + \mu_1 z_1)^{-\beta_1 - 1} z_1 - \dfrac{\lambda_2}{\phi_2 \log_e 2}\left(1 + \dfrac{\mu_2 z_2}{1 + \mu_1 z_1}\right)^{-\beta_2 - 1} \dfrac{\mu_2 z_2 z_1}{(1 + \mu_1 z_1)^2} - \kappa_1 = 0 \quad \forall \mathbf{z} \in \mathcal{Z}_1$ (57)

2) $\dfrac{\lambda_2}{\phi_2 \log_e 2}\left(1 + \dfrac{\mu_2 z_2}{1 + \mu_1 z_1}\right)^{-\beta_2 - 1} \dfrac{z_2}{1 + \mu_1 z_1} - \kappa_2 = 0 \quad \forall \mathbf{z} \in \mathcal{Z}_1$ (58)

3) $\dfrac{\lambda_1}{\phi_1 \log_e 2}\left(1 + \dfrac{\mu_1 z_1}{1 + \mu_2 z_2}\right)^{-\beta_1 - 1} \dfrac{z_1}{1 + \mu_2 z_2} - \kappa_1 = 0 \quad \forall \mathbf{z} \in \mathcal{Z}_2$ (59)

4) $-\dfrac{\lambda_1}{\phi_1 \log_e 2}\left(1 + \dfrac{\mu_1 z_1}{1 + \mu_2 z_2}\right)^{-\beta_1 - 1} \dfrac{\mu_1 z_1 z_2}{(1 + \mu_2 z_2)^2} + \dfrac{\lambda_2}{\phi_2 \log_e 2}(1 + \mu_2 z_2)^{-\beta_2 - 1} z_2 - \kappa_2 = 0 \quad \forall \mathbf{z} \in \mathcal{Z}_2$ (60)

where (57) and (58) are obtained by differentiating the Lagrangian with respect to $\mu_1$ and $\mu_2$, respectively, over $\mathbf{z} \in \mathcal{Z}_1$. Similarly, (59) and (60) are obtained by differentiating with respect to $\mu_1$ and $\mu_2$, respectively, over $\mathbf{z} \in \mathcal{Z}_2$. Due to the convexity, whenever $\mu_i, i = 1, 2$ is negative valued, we set $\mu_i = 0, i = 1, 2$. Although obtaining closed form expressions from the optimality conditions seems to be unlikely, we can gather several insights on the power control policies by analyzing the equations (57)–(60).

Let us first define $\alpha_1 = \frac{\kappa_1 \phi_1 \log_e 2}{\lambda_1}$, $\alpha_2 = \frac{\kappa_2 \phi_2 \log_e 2}{\lambda_2}$, $\alpha_{12} = \frac{\kappa_2 \phi_1 \log_e 2}{\lambda_1}$, and $\alpha_{21} = \frac{\kappa_1 \phi_2 \log_e 2}{\lambda_2}$, where $\kappa_1, \kappa_2$ are the Lagrange multipliers whose values are chosen to satisfy the average power constraint (7) with equality, and $\phi_1$ and $\phi_2$ are defined in (55) and (56). Now, consider (57) and (58). The channel state lies in $\mathcal{Z}_1$. Through a simple computation using (58), we can derive

$$\mu_2 = \frac{(1 + \mu_1 z_1)^{\frac{\beta_2}{\beta_2 + 1}}}{\alpha_2^{\frac{1}{\beta_2 + 1}} z_2^{\frac{\beta_2}{\beta_2 + 1}}} - \frac{1 + \mu_1 z_1}{z_2} \quad (61)$$

which tells us that $\mu_2 = 0$ if

$$\frac{z_2}{1+\mu_1 z_1} < \alpha_2. \tag{62}$$

If $\mu_2 = 0$, we have from (57) that

$$\frac{\lambda_1}{\phi_1 \log_e 2}(1+\mu_1 z_1)^{-\beta_1-1} z_1 - \kappa_1 = 0 \tag{63}$$

which gives us that

$$\mu_1 = \frac{1}{\alpha_1^{\frac{1}{\beta_1+1}} z_1^{\frac{\beta_1}{\beta_1+1}}} - \frac{1}{z_1} \tag{64}$$

which implies that $\mu_1 = 0$ if

$$z_1 < \alpha_1. \tag{65}$$

Now, if we substitute (61) into (57), we obtain the following additional condition for having $\mu_1 = 0$: the equation

$$\frac{z_1}{\alpha_1}(1+\mu_1 z_1)^{-(\beta_1+1)} - \frac{z_1 \alpha_2}{z_2 \alpha_{12}}\left(\left(\frac{z_2}{\alpha_2(1+\mu_1 z_1)}\right)^{\frac{1}{\beta_2+1}} - 1\right) - 1 = 0 \tag{66}$$

has a solution that returns a negative or zero value for $\mu_1$. The above discussion enables us to characterize the regions in which one user transmits while the other one is silent. We also have a closed-form formula in (64) for the optimal power adaptation policy when only one user transmits. Indeed, this is the optimal power control policy derived in [13] for a single-user system. When both users transmit, the power control policies $(\mu_1, \mu_2)$ are given directly by the non-negative solution of (57) and (58).

Note that the conditions and characterizations provided in (61)–(66) pertain to the case in which the channel state is in region $\mathcal{Z}_1$. Following a similar analysis of (59) and (60), we can obtain similar results for the cases in which the channel state is in $\mathcal{Z}_2$.

For a given partition $\{\mathcal{Z}_1, \mathcal{Z}_2\}$, the optimal power control policy can be determined numerically using the optimality conditions in (57) – (60). Additionally, the equations and inequalities in (61) through (66) can be used to guide the numerical algorithms as they specify under which conditions at most one user transmits, and provide the optimal power control policy in such cases. However, there is one difficulty. (61) – (66) depend on $\alpha_1, \alpha_2, \alpha_{12}$, and $\alpha_{21}$ which in turn depend on $\phi_1$, $\phi_2$, $\kappa_1$, and $\kappa_2$ which are in general functions of the power control policies. In such a situation, the following iterative procedure can be employed in search of the solution . We can first choose certain values for $\phi_1$, $\phi_2$, $\kappa_1$, and $\kappa_2$, and then determine the optimal power allocation policies for these selected values. Subsequently, we can check whether the obtained policy satisfies the average power constraint with equality. This enables us to determine if the selected $\kappa_1$ and $\kappa_2$ values are accurate. We can also compute $\phi_1$ and $\phi_2$ using the obtained policy and see if they agree with the initial values of $\phi_1$ and $\phi_2$. If there is no sufficient match or if the power constraint is not satisfied with equality, then we update the values of $\phi_1$, $\phi_2$, $\kappa_1$, and $\kappa_2$, and reiterate the search of the optimal policy.

With this insight, we propose the following algorithm that can be used to determine the optimal power allocated to each channel state:

POWER CONTROL ALGORITHM

1. Given $\lambda_1, \lambda_2$, the partition $\mathcal{Z}$, initialize $\phi_1, \phi_2$;
2. Initialize $\kappa_1$ and $\kappa_2$;
3. Determine $\alpha_1 = \frac{\kappa_1 \phi_1 \log_e 2}{\lambda_1}$, $\alpha_2 = \frac{\kappa_2 \phi_2 \log_e 2}{\lambda_2}$, $\alpha_{12} = \frac{\kappa_2 \phi_1 \log_e 2}{\lambda_1}$, $\alpha_{21} = \frac{\kappa_1 \phi_2 \log_e 2}{\lambda_2}$;
4. **if** $\mathbf{z} \in \mathcal{Z}_1$
5.    **then if** $z_2 > \alpha_2$
6.       **then** $\mu_2 = \frac{1}{\alpha_2^{\frac{1}{\beta_2+1}} z_2^{\frac{\beta_2}{\beta_2+1}}} - \frac{1}{z_2}$;
7.          **if** $\frac{z_1}{\alpha_1}(1+\mu_1 z_1)^{-(\beta_1+1)} - \frac{z_1 \alpha_2}{z_2 \alpha_{21}}\left(\left(\frac{z_2}{\alpha_2(1+\mu_1 z_1)}\right)^{\frac{1}{\beta_2+1}} - 1\right) - 1 = 0$ returns nonpositive $\mu_1$
8.             **then** $\mu_1 = 0$;
9.             **else if** $\frac{z_2}{\alpha_2} < \left(\frac{z_1}{\alpha_1}\right)^{\frac{1}{\beta_1+1}}$
10.                 **then** $\mu_2 = 0$, $\mu_1 = \left[\frac{1}{\alpha_1^{\frac{1}{\beta_1+1}} z_1^{\frac{\beta_1}{\beta_1+1}}} - \frac{1}{z_1}\right]^+$;
11.             **else** Compute $\mu_1, \mu_2$ from (57) and (58);
12.       **else** $\mu_2 = 0$, $\mu_1 = \left[\frac{1}{\alpha_1^{\frac{1}{\beta_1+1}} z_1^{\frac{\beta_1}{\beta_1+1}}} - \frac{1}{z_1}\right]^+$;
13. **if** $\mathbf{z} \in \mathcal{Z}_2$
14.    **then if** $z_1 > \alpha_1$
15.       **then** $\mu_1 = \frac{1}{\alpha_1^{\frac{1}{\beta_1+1}} z_1^{\frac{\beta_1}{\beta_1+1}}} - \frac{1}{z_1}$;
16.          **if** $\frac{z_2}{\alpha_2}(1+\mu_2 z_2)^{-(\beta_2+1)} - \frac{z_2 \alpha_1}{z_1 \alpha_{21}}\left(\left(\frac{z_1}{\alpha_1(1+\mu_2 z_2)}\right)^{\frac{1}{\beta_1+1}} - 1\right) - 1 = 0$ returns nonpositive $\mu_2$
17.             **then** $\mu_2 = 0$;
18.             **else if** $\frac{z_1}{\alpha_1} < \left(\frac{z_2}{\alpha_2}\right)^{\frac{1}{\beta_2+1}}$
19.                 **then** $\mu_1 = 0$, $\mu_2 = \left[\frac{1}{\alpha_2^{\frac{1}{\beta_2+1}} z_2^{\frac{\beta_2}{\beta_2+1}}} - \frac{1}{z_2}\right]^+$;
20.             **else** Compute $\mu_1, \mu_2$ from (59) and (60);
21.    **else** $\mu_1 = 0$, $\mu_2 = \left[\frac{1}{\alpha_2^{\frac{1}{\beta_2+1}} z_2^{\frac{\beta_2}{\beta_2+1}}} - \frac{1}{z_2}\right]^+$;
22. Check if the obtained power control policies $\mu_1$ and $\mu_2$ satisfy the power constraint with equality;
23. **if** not satisfied with equality
24.    **then** update the values of $\kappa_1$ and $\kappa_2$ and return to Step 3;
25.    **else** move to Step 26;
26. Evaluate $\phi_1$ and $\phi_2$ with the obtained power control policies;
27. Check if the new values of $\phi_1$ and $\phi_2$ agree (up to a certain margin) with those used in Step 3;
28. **if** do not agree
29.    **then** update the values of $\phi_1$ and $\phi_2$ and return to Step 2;
30.    **else** declare the obtained power allocation policies $\mu_1$ and $\mu_2$ as the optimal ones.

Note that we above have not specified how the values of $\kappa_1, \kappa_2, \phi_1$, and $\phi_2$ are updated for each iteration in order to keep the algorithm generic. In our numerical computations, we have updated $\kappa_1$ and $\kappa_2$ using the bisection search algorithm. The values of $\phi_1$ and $\phi_2$ are updated in Step 29 of the algorithm by assigning them the values evaluated in Step 26. Hence, the most recent values are carried over to the new iteration.

In Fig. 5, we plot the optimal power allocation policies $\mu_1$ and $\mu_2$ as functions of channel fading states $z_1$ and $z_2$. We assume that $\theta_1 = \theta_2 = 0.01$, $\text{SNR}_1 = \text{SNR}_2 = 0$ dB, and $\lambda_1 = \lambda_2 = 0.5$. We consider the partition specified by the suboptimal decoding order given in (42). Hence, since we have $\lambda_1 = \lambda_2 = 0.5$, decoding orders (1,2) and (2,1) are used when $z_2 < z_1$ and $z_2 > z_1$, respectively. Under these assumptions, we computed the optimal values as $\kappa_1^* = 0.0470$, $\kappa_2^* = 0.0462$, $\phi_1^* = 0.5550$, and $\phi_2^* = 0.5538$. In the figure, we observe that each user, not surprisingly, allocates most of its power to the regions in which it is decoded last and hence does not experience interference. However, due to the introduction of QoS constraints, we also note that each user also allocates certain power to the cases in which it is decoded first. This is performed in order to continue transmission and avoid buffer overflows.

So far, we have assumed that the partition $\mathcal{Z}$ is given. The optimal partition $\mathcal{Z}$ that maximizes the weighted sum-rate can be derived through the following optimization similarly as in [21]:

$$\mathsf{C}^* = \sup_{\mathcal{Z}} \lambda_1 \mathsf{C}_1(\mu, \mathcal{Z}) + \lambda_2 \mathsf{C}_2(\mu, \mathcal{Z}) \tag{67}$$

where $\mathsf{C}^*$ is the optimal weighted sum value for given pair of $(\lambda_1, \lambda_2)$, and $\mu = (\mu_1, \mu_2)$ are the optimal power control policies for given $\mathcal{Z}$.

## VI. Conclusion

In this paper, we have studied the achievable throughput regions in multiple access fading channels when users operate under QoS constraints. We have assumed that both the transmitters and the receiver have perfect CSI. We have employed the effective capacity as a measure of the throughput under buffer constraints. We have defined the effective capacity region and shown its convexity. We have considered different transmission and reception scenarios e.g, superposition coding, different strategies for the decoding order, and TDMA. Under the assumption that no power control is employed by the transmitters, we have analyzed the performances of fixed and variable decoding order strategies. We have characterized the throughput region and determined the points on its boundary for fixed decoding order. For the case of two users with the same QoS constraints, we have derived the optimal strategy for varying the decoding order. Varying the decoding order is shown to significantly increase the achievable rate region. We have also proposed a simpler suboptimal decoding rule which can almost perfectly match the optimal throughput region. We have also studied the performance of orthogonal transmission strategies by considering TDMA. In the numerical results, we have demonstrated that TDMA can perform better than superposition coding with fixed decoding order for certain QoS constraints. More specifically, we have noted that TDMA can support arrival rate pairs that are strictly outside the region achieved when fixed decoding order is employed at the receiver. We have also observed that the performance of TDMA approaches that of the optimal strategy of superposition coding with variable decoding order as $\theta$ increases (i.e., as the QoS constraints become more stringent). In the second part of the paper, we have incorporated power adaptation strategies into the model. For a given fixed decoding order at the receiver, we have identified the optimal power control policies. For cases in which a variable decoding order strategy is adopted by the transmitter, we have obtained the conditions that the optimal strategies should satisfy and described an algorithm to achieve these optimal schemes.

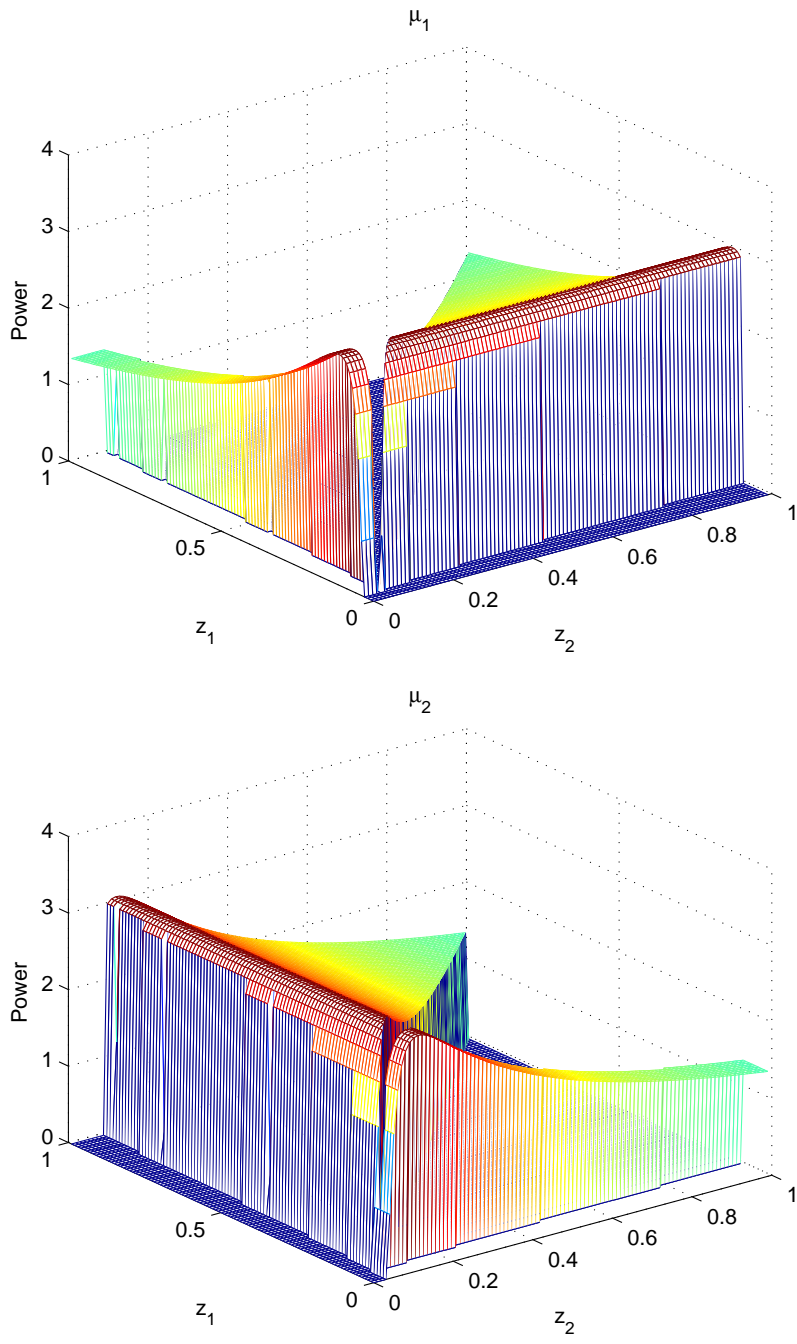

Fig. 5. The optimal power control policies $\mu_1$ and $\mu_1$ of users 1 and 2, respectively, as a function of $(z_1, z_2)$. $\lambda_1 = 0.5$, $\lambda_2 = 0.5$.

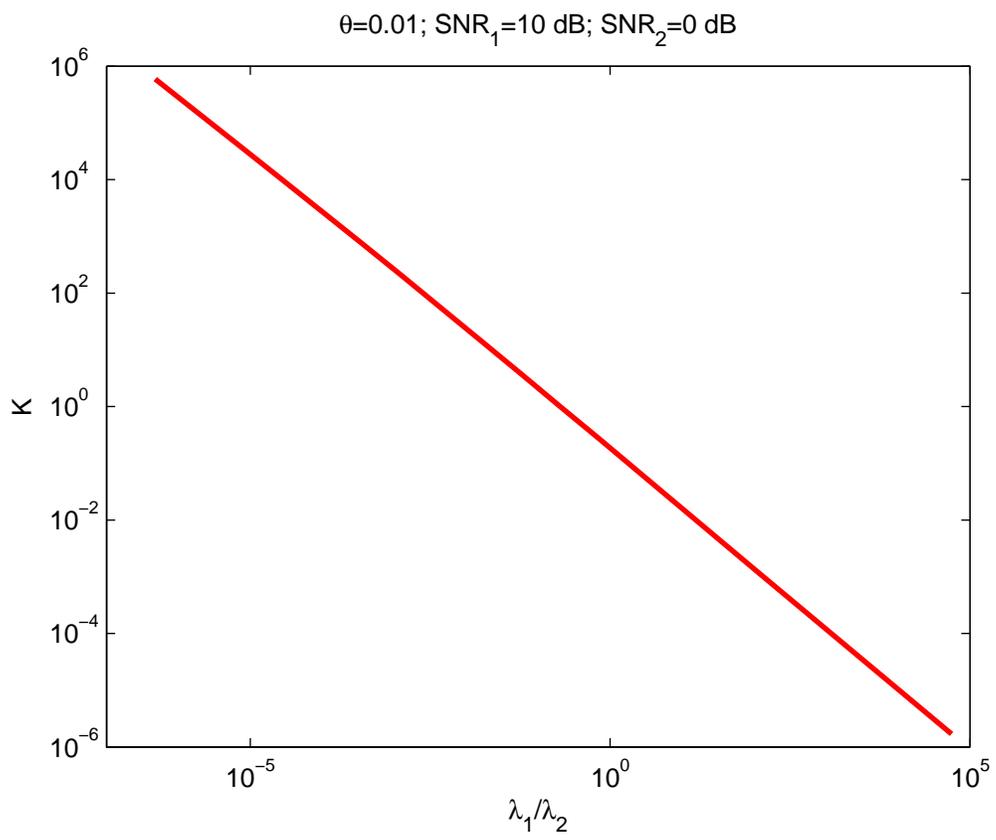